\documentclass[twocolumn,showpacs,amsmath,amssymb]{revtex4-1}
\usepackage{graphicx}
\begin{document}
\title{\Large \bf Decoupling and Reduction in  Chern-Simons Modified Gravity}
\author{\large Haji Ahmedov and Alikram N. Aliev}
\address{Feza G\"ursey Institute, \c Cengelk\" oy, 34684   Istanbul, Turkey}
\date{\today}

\begin{abstract}

We show that for four-dimensional spacetimes with a non-null hypersurface orthogonal Killing vector and for a Chern-Simons (CS) background (non-dynamical) scalar field, which is constant along the Killing vector, the source-free  equations of CS modified gravity  decouple into their Einstein and Cotton constituents. Thus, the model supports only  general relativity solutions. We also show that, when the cosmological constant vanishes and the gradient of the CS scalar field is parallel to the non-null hypersurface orthogonal Killing vector of constant length, CS modified gravity  reduces to topologically massive gravity in three dimensions. Meanwhile, with the cosmological constant such a reduction requires an appropriate source term  for CS modified gravity.

\end{abstract}


\maketitle

\section{Introduction}

Chern-Simons modified gravity is an intriguing extension of general relativity, which amounts to complementing the Hilbert action
with a parity-violating Pontryagin density coupled to a CS scalar field \cite{jackiw}.  The Pontryagin density is defined as a contraction of the Riemann curvature tensor with its dual and the CS scalar field can be viewed as either a prescribed background quantity or as an evolving field. Below we shall consider only the case of a prescribed CS scalar field within the non-dynamical formulation of the theory. One of the attractive features of CS modified gravity is its emergence within predictive frameworks of more fundamental physical theories. For instance, the low-energy limit of string theory comprises  general relativity with a parity-violating correction term, that is nothing but the Pontryagin density. This term is crucial  for canceling gravitational anomaly in string theory through Green-Schwarz mechanism \cite{green}. The Pontryagin density, as an anomaly-canceling term, also arises in particle physics and in the context of loop quantum gravity (see a recent review  \cite{ay1} for further details).

It is interesting to note that the authors of work \cite{jackiw} arrived at CS modified gravity by  uplifting the three-dimensional gravitational Chern-Simons term to four-dimensional spacetime. Three-dimensional gravity with the Chern-Simons term  known as topologically massive gravity (TMG) cures dynamically trivial nature of Einstein's gravity in three dimensions  \cite{djt}. Embedding the  Chern-Simons term of TMG into four-dimensional spacetime results in a topological current, whose divergence is given by the Pontryagin density. Subsequently, the associated total gravitational action with  an appropriate embedding coordinate (with the gradient of the CS scalar field) looks precisely the same as that obtained for CS modified gravity in the low-energy limit of string theory \cite{td}. Meanwhile, the   field equations  consist  of two parts: the Einstein part and the Chern-Simons part, that is given by a symmetric and traceless Cotton-type tensor. In the context of uplifting of TMG to four dimensions, the latter tensor arises as an analogue of the three-dimensional Cotton tensor. Thus, in a certain sense,  CS modified gravity can be thought of as a four-dimensional ``counterpart" of TMG.

Motivated by this fact, in this Letter we  present a decoupling theorem for CS modified gravity. This is  an extension of a similar theorem  for TMG, earlier proved in \cite{an} (see also a recent work \cite{df}), to four dimensions. Namely, we  prove that for four-dimensional spacetimes admitting a nun-null hypersurface orthogonal Killing vector and for a CS scalar field being constant along the Killing vector (the Lie derivative of the CS scalar field vanishes along the Killing vector), source-free CS modified gravity decouples into its Einstein and Cotton parts. We also present a reduction theorem, showing that for zero cosmological constant and for the gradient of the CS scalar field being parallel to the non-null hypersurface orthogonal Killing vector of constant length, CS modified gravity reduces to TMG in three dimensions. Finally, we show that in the case of nonvanishing cosmological constant, such a reduction requires a specific source term in the field equations of  CS modified gravity.  This source is determined by the cosmological constant.

\section{Basics of CS Modified Gravity}

CS modified gravity is an extension of four-dimensional general relativity by adding to the usual Hilbert action a gravitational parity-violating term, the Pontryagin density \cite{jackiw}. The full action including a cosmological term has the form
\begin{eqnarray}
S &=& \frac{1}{16\pi G} \int d^4x \sqrt{-g}\left(R - 2 \Lambda + \frac{1}{4} \, \vartheta \,^\ast{R} R\right)\,,
\label{csaction}
\end{eqnarray}
where $ R $ is the Ricci scalar, $ \Lambda $ is the cosmological constant,  $ \vartheta \, $ is the CS coupling  scalar field and  the
Pontryagin density
\begin{eqnarray}
^\ast{R} R &=& {^\ast{R}^{\mu}_{~\,\nu}}^{\lambda \tau} R^{\nu}_{~\,\mu \lambda \tau}\,.
\label{pontryagin}
\end{eqnarray}
The dual of the Riemann curvature tensor is  defined as
\begin{eqnarray}
{^\ast{R}^{\mu}_{~\,\nu}}^{\lambda \tau}&=& \frac{1}{2} \, \epsilon^{\,\lambda \tau \rho \sigma } R^{\mu}_{~\,\nu \rho \sigma}\,,
\label{dual}
\end{eqnarray}
where $ \epsilon^{\,\lambda \tau \rho \sigma } $ is the totally antisymmetric Levi-Civita tensor $ (\epsilon^{\,\lambda \tau \rho \sigma}= \frac{1}{\sqrt{-g}}\,\varepsilon^{\,\lambda \tau \rho \sigma},~\varepsilon^{0123}=1 ).$

It is curious that the Pontryagin density can also be expressed in terms of the  divergence of a four-dimensional topological current. We have
\begin{eqnarray}
^\ast{R} R &= & 2 \nabla_{\mu}K^{\mu}\,,
\label{divcurrent}
\end{eqnarray}
where the topological current is given by
\begin{eqnarray}
K^{\mu} &= & \epsilon^{\,\mu \nu \alpha \beta} \Gamma^{\lambda}_{\nu \tau}\left(\partial_{\alpha}\Gamma^{\tau}_{\beta \lambda}
+\frac{2}{3}\,\Gamma^{\tau}_{\alpha \sigma}\Gamma^{\sigma}_{\beta \lambda}\right)\,,
\label{current}
\end{eqnarray}
and the quantities $ \Gamma^{\mu}_{\alpha \beta} $  are the usual Christoffel symbols. It is easy to check that for the CS scalar field $ \vartheta = const $, the Pontryagin  density does not contribute to the field equations and the theory reduces to general relativity.

The variation of action (\ref{csaction}) with respect to the spacetime metric leads to the field equations of CS modified gravity in the form
\begin{eqnarray}
E_{\mu\nu} + C_{\mu\nu} &= & 0\,,
\label{fieldeqs}
\end{eqnarray}
where
\begin{eqnarray}
E_{\mu\nu} &= & R_{\mu\nu} - \frac{1}{2}\, R g_{\mu\nu} + \Lambda g_{\mu\nu} \,,
\label{Etensor}
\end{eqnarray}
$ R_{\mu\nu} $ is the Ricci tensor and $ C_{\mu\nu} $ is a symmetric and traceless tensor \cite{jackiw}. We have
\begin{eqnarray}
C^{\mu\nu} &= & \vartheta_{\alpha}\,\epsilon^{\,\alpha \beta \gamma(\mu} R^{ \nu)}_{ \beta \,;\gamma}
+\vartheta_{\alpha \beta}\,^\ast{R}^{\beta (\mu \nu)\alpha}\,\,,
\label{cotton4d}
\end{eqnarray}
where the semicolon stands for covariant differentiation and
\begin{eqnarray}
\vartheta_{\alpha}& = & \vartheta_{;\alpha} ~~~~~ \vartheta_{\alpha \beta}=  \vartheta_{\alpha ; \beta}\,.
\label{embcoord}
\end{eqnarray}
As we have mentioned above, in the context of CS modified gravity the tensor in (\ref{cotton4d}) can be thought of  as a four-dimensional ``image" of the usual Cotton tensor in three dimensions. Therefore, for the sake of convenience,   we shall simply call it the four-dimensional ``Cotton" tensor (see also Ref. \cite{jackiw}).

It is important to note  that the four-dimensional Cotton tensor has nonvanishing divergence, unlike the three-dimensional Cotton tensor. It is straightforward  to show that
\begin{eqnarray}
C^{\mu\nu}_{~~;\nu} &= & -\frac{1}{8}\,\vartheta^{\mu}\, ^\ast{R} R\,.
\label{cotdiv}
\end{eqnarray}
On the other hand, the contracted Bianchi identities applied to the field equations in (\ref{fieldeqs}) require the Pontryagin density to vanish. Thus, we have the constraint equation
\begin{eqnarray}
 ^\ast{R} R &= & 0\,.
\label{pconstraint}
\end{eqnarray}
We note that this  equation naturally arises when  varying  action  (\ref{csaction}) with respect to CS scalar field. This constraint is only a necessary condition  for a spacetime to be a solution to CS modified gravity, as it does not mean the vanishing of the Cotton tensor itself. It is clear that this condition will  restrict the class of solutions to CS modified gravity.

\section{Decoupling}

In this section, we shall show that another restriction for the solution space of CS modified gravity is given by  decoupling of the field equations (\ref{fieldeqs}) into their Einstein and Cotton parts. This is  expressed in the following theorem:

{\it If a four-dimensional spacetime admits a non-null hypersurface orthogonal Killing vector and the CS coupling scalar field is constant along the Killing vector, then the field equations of CS modified gravity  with a cosmological constant decouple into their Einstein and Cotton constituents, thus supporting only  general relativity solutions}.

Since it is supposed that the CS coupling scalar field is constant along the hypersurface orthogonal Killing vector, we have
\begin{eqnarray}
\pounds_{\xi} \vartheta  & = &  \left(\vartheta_{\alpha}\xi^{\alpha}\right) = 0\,,
\label{orthscalar}
\end{eqnarray}
where $ \pounds_{\xi} $ is the Lie derivative along the Killing vector $ \xi  $. In other words, the vectors $\, \xi_{\alpha} $ and $ \vartheta_{\alpha} $ are orthogonal to each other. Before proceeding with the proof of the theorem we need to explore some consequences of  the hypersurface orthogonal  Killing vector. The defining equations for this vector are given by
\begin{equation}
\xi_{(\mu ;\nu)} = 0\,,~~~~~~\xi_{[\mu ; \nu} \,\xi_{\lambda]} = 0\,,
\label{killingh}
\end{equation}
where the second equation is the hypersurface orthogonality condition. Here  and in  what follows, we  use round and square brackets to denote symmetrization and antisymmetrization of indices, respectively.

Taking the covariant derivative of this condition  and using the fact that for any Killing vector \cite{wald}
\begin{equation}
\xi_{\mu; \nu ; \lambda} =  \xi_{\tau}  R^{\tau}\,_{\lambda \nu \mu}\,,
\label{riemann2}
\end{equation}
we obtain \cite{an}
\begin{equation}
\xi_{\mu}  R^{\mu}\,_{\nu[\alpha \beta}\, \xi_{\gamma]} +  \xi_{[\beta ;\alpha}\, \xi_{\gamma];\nu} =  0\,.
\label{hyperder}
\end{equation}
On the other hand, contracting the hypersurface orthogonality condition in  (\ref{killingh}) with the Killing vector $ \xi^{\lambda} $, we have \begin{equation}
\xi_{\mu ;\nu} = \xi_{[\mu} \eta_{\nu]}\,,
\label{cdkilling}
\end{equation}
where the vector $ \eta _{\mu}$ is given by
\begin{equation}
\eta_\mu= \xi^{-2}\, \xi^2_{\,;\mu}\,
\label{eta}
\end{equation}
and $ \xi^{2} $ is the  square of the length  of $ \xi_\mu $ \cite{wald}.  Next, using equation (\ref{cdkilling}) in equation (\ref{hyperder}) we find that the latter  reduces to the form
\begin{equation}
\xi_{\mu} R^{\mu}\,_{\nu[\alpha \beta}\, \xi_{\gamma]}  =  0\,.
\label{hyperder1}
\end{equation}
This equation, when contracted over the indices $\nu $ and $ \alpha $, yields \cite{carter}
\begin{equation}
\xi^{\lambda} \, R_{\lambda[\mu}\,\xi_{\nu]}=0\,,
\label{rel1}
\end{equation}
which, in turn, implies that
\begin{equation}
\xi^{\mu}  R_{\mu \nu} = \omega \xi_{\nu}\,,
\label{rel2}
\end{equation}
where $ \omega $ is a scalar function.

For further use, we also need to know the covariant derivative of equation (\ref{cdkilling}). After some manipulations, we have
\begin{eqnarray}
\xi_{\mu ;\nu;\lambda}&=& \xi_{[\mu}\eta_{\nu];\lambda}+\frac{1}{2}\, \eta_{\lambda}\xi_{[\mu}\eta_{\nu]}\,.
\label{rel3}
\end{eqnarray}
Next, using the fact  that the Lie derivative of the Ricci  tensor  along the Killing vector vanishes,
\begin{equation}
\pounds_{\xi} R_{\mu\nu} = R_{\mu \nu; \lambda} \, \xi^{\lambda}
 + R^{\lambda}_{\;\mu} \,\xi_{\lambda ;\nu}  + R^{\lambda}_{\; \nu} \,\xi_{\lambda  ;\mu} = 0\,,
\label{lie2}
\end{equation}
we obtain the relation
\begin{equation}
R_{\mu \nu; \lambda} \, \xi^{\lambda}=\eta^{\lambda}R_{\lambda(\mu} \xi_{\nu)}- \omega\,\xi_{(\mu}\eta_{\nu)}\,,
\label{liecom}
\end{equation}
where we have also used  equations (\ref{cdkilling}) and (\ref{rel2}).

We are now ready to  turn to the proof of the theorem, which will be based on exploration of  all possible  projections of  tensors (\ref{Etensor}) and (\ref{cotton4d}) in directions parallel and orthogonal to the Killing vector.

For an arbitrary  tensor field  $ A $  we shall use the notations
\begin{eqnarray}
A_{**} & = & A_{\mu \nu} \xi^{\mu} \xi^{\nu}\,,~~~~\overline{A}_{\lambda \sigma}=A_{\mu\nu}\, h_{\;\lambda}^{\mu} h_{\;\sigma}^{\nu}\,,\nonumber \\ [2mm] &&
\overline{A}_{* \lambda}= A_{\mu\nu} \xi^{\mu} h_{\;\lambda}^{\nu}\,,
\label{parapermix}
\end{eqnarray}
for the parallel, orthogonal and mixed projections, respectively. Here the projection operator $ h_{\;\nu}^{\mu} $  onto the subspace orthogonal to the Killing vector is given by
\begin{eqnarray}
h_{\;\nu}^{\mu} &=& \delta_{\;\nu}^{\mu}
    - \frac{\xi^{\mu} \xi_{\nu}} {\xi^2} \,,~~~~~
h_{\;\lambda}^{\mu} \, h_{\;\nu}^{\lambda}  = h_{\;\nu}^{\mu}\,, ~~~~~ \label{proj1}
\end{eqnarray}
and
\begin{eqnarray}
h_{\;\nu}^{\mu} \,\xi^\nu &=& 0\,.
\label{proj2}
\end{eqnarray}
For convenience, we shall first show that
\begin{eqnarray}
C_{**} & = & C_{\mu \nu} \xi^{\mu} \xi^{\nu}  \equiv  0\,.
\label{paralcot1}
\end{eqnarray}
Contracting equation (\ref{cotton4d}) with the Killing vectors $\xi_{\mu}$ and $  \xi_{\nu} $ and taking into account equations (\ref{dual}) and  (\ref{riemann2}) we  obtain, after some rearrangements, the expression
\begin{eqnarray}
C_{**}& = & \vartheta_{\alpha}\,\varepsilon^{\,\alpha \beta \gamma \mu}\,\xi_{\mu} \left[\left(R^{\nu}_{\beta} \,\xi_\nu \right)_{;\gamma} - R^{\nu}_{\beta} \,\xi_{\nu ;\gamma}\right] \nonumber \\ [2mm] &&
 -\frac{1}{2}\, \vartheta_{\alpha}^{\; \beta} \,\varepsilon^{\,\nu \alpha \gamma \tau}\,\xi_{\nu} \,\xi_{\tau ;\gamma ; \beta}\,.
\label{paralcot2}
\end{eqnarray}
Next, substituting equations (\ref{cdkilling}) and  (\ref{rel2}) in the first line and equation (\ref{rel3}) in the last line of this equation, we  have
\begin{eqnarray}
C_{**}& = & \vartheta_{\alpha}\,\varepsilon^{\,\alpha \beta \gamma \mu}\,\xi_{\mu} \left[\omega_{;\gamma}\xi_{\beta}-\frac{1}{2}\,\xi_{\gamma}\left(\omega \eta_{\beta} - R^{\nu}_{\beta}\eta_{\nu}\right)\right]
\nonumber \\ [2mm] &&
 -\frac{1}{2}\, \vartheta_{\alpha}^{\; \beta} \,\varepsilon^{\,\nu \alpha \gamma \tau}\,\xi_{\nu} \left(
\xi_{[\tau}\eta_{\gamma];\beta}+\frac{1}{2}\, \eta_{\beta}
\xi_{[\tau}\eta_{\gamma]}\right)\,.
 \label{paralcot3}
\end{eqnarray}
We see that each term in this expression vanishes identically due to the contraction of a symmetric pair of the Killing vectors with the Levi-Civita tensor. Thus, we arrive at identity (\ref{paralcot1}).  With this identity, from the field equations  (\ref{fieldeqs}) it immediately follows that
\begin{eqnarray}
E_{**} & = & E_{\mu \nu} \xi^{\mu} \xi^{\nu} =  0\,.
\label{paralE}
\end{eqnarray}
Contracting  now equation (\ref{rel2}) with the projection operator in (\ref{proj1}) and taking into account equation (\ref{proj2}), we obtain the identity
\begin{eqnarray}
\overline{R}_{*\lambda}&=&  R_{\mu \nu}\xi^{\mu}h^{\nu}_{\;\lambda} \equiv  0\,.
\label{parperric}
\end{eqnarray}
With this in mind, from equation (\ref{Etensor}) it is easy to see that
\begin{eqnarray}
\overline{E}_{*\lambda}&=&  E_{\mu \nu}\xi^{\mu}h^{\nu}_{\;\lambda} \equiv  0\,,
\label{parperE}
\end{eqnarray}
which also implies
\begin{eqnarray}
\overline{C}_{*\lambda}&=&  C_{\mu \nu}\xi^{\mu}h^{\nu}_{\;\lambda} =  0\,,
\label{parpercot}
\end{eqnarray}
as a consequence of the field equations (\ref{fieldeqs}).

We recall that separation into  the pairs of equations (\ref{paralcot1}),  (\ref{paralE}) and (\ref{parperE}), (\ref{parpercot}) occurs regardless  of the form of CS scalar field. However, as we shall see below, this is not the case for the remaining orthogonal projections of the Einstein and Cotton parts, for which such a separation occurs only for the  CS  scalar field obeying the relation (\ref{orthscalar}).

To proceed further, we first note that from the obvious relation
\begin{equation}
h_{\;\alpha}^{\mu}h_{\;\beta}^{\nu} h_{\;\gamma}^{\lambda}h_{\;\sigma}^{\tau} \epsilon^{\,\alpha \beta \gamma \sigma}=0\,,
\label{perlevi}
\end{equation}
it follows that
\begin{equation}
\epsilon^{\,\mu \nu \lambda \tau}= \frac{\xi_{\gamma}}{\xi^2}\,\left(\xi^{\mu}\epsilon^{\,\gamma \nu \lambda \tau}+ \xi^{\nu}\epsilon^{\,\mu \gamma  \lambda \tau} + \xi^{\lambda}\epsilon^{\,\mu \nu \gamma \tau}+\xi^{\tau}\epsilon^{\,\mu \nu \lambda \gamma} \right),
\label{leviexpan}
\end{equation}
where we have used equation (\ref{proj1}). We shall now calculate
the orthogonal projection of the Cotton tensor, defining it as
\begin{eqnarray}
\overline{C}^{\lambda \sigma}& = & C^{\mu\nu} h^{\lambda}_{\;\mu} h^{\sigma}_{\nu} \,.
\label{cotorthog1}
\end{eqnarray}
With equations (\ref{proj2}) and  (\ref{leviexpan}) in mind, we substitute into this expression  the explicit form of the Cotton tensor (\ref{cotton4d}). As a consequence, we find that
\begin{eqnarray}
\overline{C}^{\lambda \sigma} & = & \frac{h^{(\lambda}_{\;\mu} h^{\sigma)}_{\;\nu} \, \xi_{\tau}}{\xi^2}\,\left\{\vartheta_{\alpha}\left(\xi^{\beta}\epsilon^{\,\alpha \tau \gamma \mu}+ \xi^{\gamma}\epsilon^{\,\alpha \beta \tau \mu}\right) R^{\nu}_{\beta ;\gamma}\nonumber
\right. \\[2mm]  & & \left.
~~~~~~~~~~~~~+(\vartheta_{\alpha}\xi^{\alpha})\,\epsilon^{\, \tau \beta \gamma \mu} R^{\nu}_{\beta ;\gamma}\nonumber
\right. \\[2mm]  & & \left.
+ \frac{1}{2}\,\vartheta_{\alpha \beta}\left(\xi^{\alpha} \epsilon^{\, \nu \tau \rho \gamma} + 2\xi^{\rho}\epsilon^{\, \nu \alpha \tau \gamma}\right){R^{\beta \mu}}_{\rho \gamma}\right\}\,.
\label{cotperper1}
\end{eqnarray}
Substituting in the first line of this equation the relation
\begin{eqnarray}
R^{\nu}_{\beta ; \gamma}\xi^{\beta}& = & (\omega \xi^{\nu})_{;\gamma} -R^{\nu}_{\beta}\, \xi^{\beta}_{\,\;;\gamma}
\label{use1}
\end{eqnarray}
as well as equation (\ref{liecom}) and taking  into account (\ref{cdkilling}), it is easy to see that each term in this line vanishes identically due to either equation  (\ref{proj2}) or the contraction of a symmetric combination of the Killing vectors with the Levi-Civita tensor. Meanwhile, using equation (\ref{riemann2}) in the last line,  we   transform  equation (\ref{cotperper1}) into the form
\begin{eqnarray}
\overline{C}^{\lambda \sigma} & = & \frac{h^{(\lambda}_{\;\mu} h^{\sigma)}_{\;\nu} \, \xi_{\tau}}{\xi^2}\,\left\{(\vartheta_{\alpha}\xi^{\alpha})\,\epsilon^{\, \tau \beta \gamma \mu} R^{\nu}_{\beta ;\gamma}+ \frac{1}{2}\,\vartheta_{\alpha \beta}\xi^{\alpha}
\nonumber
\right. \\[2mm]  & & \left.
\left[\epsilon^{\, \nu \tau \rho \gamma}{R^{\beta \mu}}_{\rho \gamma}+ \epsilon^{\, \nu \beta \tau \gamma}(\eta^{\mu}_{\,;\gamma}+\frac{1}{2}\,\eta^{\mu}\eta_{\gamma})\right]\right\}.
\label{cotperper2}
\end{eqnarray}
For some purposes, it may be  useful to make a further simplification of this expression. To achieve this goal, we  introduce the derivative operator $ D_{\mu} $  with  respect to the three-dimensional metric  $ h_{\mu\nu} $ in the usual way \cite{wald}
\begin{eqnarray}
D_{\nu} V_{\mu} & = & h^{\lambda}_{\;\mu}  h^{\sigma}_{\;\nu} \, V_{\lambda ; \sigma}\,.
\label{3doper}
\end{eqnarray}
Using this in the definition of the Riemann tensor
\begin{eqnarray}
^{(3)}{R^{\mu}}_{\nu \lambda \tau} V_{\mu} & = & 2 D_{[\tau} D_{\lambda]} V_{\nu }
\label{defcur}
\end{eqnarray}
and taking into account the fact that in our case the second fundamental form  vanishes, i.e.
\begin{eqnarray}
h^{\lambda}_{\;\mu}  h^{\sigma}_{\;\nu} \xi_{\lambda ; \sigma} & = & 0 \,, \label{useful2}
\end{eqnarray}
it is easy to see that
\begin{eqnarray}
^{(3)}R_{\mu \nu \lambda \tau} & = & h^{\alpha}_{\;\mu}h^{\beta}_{\;\nu} h^{\rho}_{\;\lambda} h^{\sigma}_{\;\tau} R_{\alpha\beta\rho\sigma}\,
\label{cur3to4}
\end{eqnarray}
as a consequence of Gauss-Codacci equation \cite{wald}.
From this equation, we also find that
\begin{eqnarray}
^{(3)}R_{\mu \nu} & = & h^{\alpha}_{\;\mu}h^{\beta}_{\;\nu} \left[ R_{\alpha\beta}+ \frac{1}{2} (\eta_{\alpha ;\beta}+\frac{1}{2}\,\eta_{\alpha}\eta_{\beta})\right].
\label{ricci3to4}
\end{eqnarray}
In obtaining this expression we have used  equations (\ref{riemann2}) and (\ref{cdkilling}). Similarly, for the three-dimensional scalar curvature we have
\begin{eqnarray}
^{(3)}R & = &  R+ \eta^{\alpha}_{\;;\alpha}
\label{ricscalar3to4}
\end{eqnarray}
We also recall that
\begin{eqnarray}
^{(3)}R_{\mu \nu \lambda \tau} & = & 2\left(^{(3)}R_{\mu [\lambda}h_{\tau]\nu} - \,^{(3)}R_{\nu [\lambda} h_{\tau]\mu}\nonumber
\right. \\[2mm]  & & \left.
-\frac{^{(3)}R}{2}\, h_{\mu[\lambda}h_{\tau]\nu}\right).
\label{curcur3to4}
\end{eqnarray}
Combining equations (\ref{cur3to4})-(\ref{curcur3to4}) and using the result in equation (\ref{cotperper2}) we obtain, after some straightforward  calculations,  the following  expression for the orthogonal projection of the Cotton tensor
\begin{eqnarray}
\overline{C}^{\,\lambda \sigma} & = & \frac{\xi_{\tau}}{\xi^2}\,\left[
(\vartheta_{\alpha}\xi^{\alpha})\,\epsilon^{\,\tau \beta \gamma (\lambda}D_{\gamma}\overline{R}^{\,\sigma)}_{\beta}- \vartheta_{\alpha \beta}\xi^{\alpha}
\epsilon^{\,\tau \beta \gamma (\lambda}\overline{R}^{\,\sigma)}_{\gamma} \right], \nonumber\\
\label{finalcotperp}
\end{eqnarray}
where
\begin{eqnarray}
\vartheta_{\alpha \beta}\xi^{\alpha}& = & \left(\vartheta_{\alpha}\xi^{\alpha}\right)_{; \beta}+\frac{1}{2}\left[\left(\vartheta_{\alpha}\eta^{\alpha}\right) \xi_{\beta}- \left(\vartheta_{\alpha}\xi^{\alpha}\right) \eta_{\beta}\right].
\label{vab}
\end{eqnarray}
We note that the latter expression is obtained by using equations (\ref{embcoord}) and (\ref{cdkilling}). Clearly,  the expression in (\ref{finalcotperp}) does not vanish in general. However, with  condition (\ref{orthscalar}) it vanishes identically. Thus, we have
\begin{eqnarray}
\overline{C}^{\lambda \sigma}& = & C^{\mu\nu} h^{\lambda}_{\;\mu} h^{\sigma}_{\nu} \equiv  0\,,
\label{cotorthogr}
\end{eqnarray}
which, by means of field equations (\ref{fieldeqs}), results in
\begin{eqnarray}
\overline{E}^{\lambda \sigma}& = & E^{\mu\nu} h^{\lambda}_{\;\mu} h^{\sigma}_{\nu} = 0\,.
\label{orthogrE}
\end{eqnarray}
These equations together with those given in  (\ref{paralcot1}),  (\ref{paralE}) and (\ref{parperE}), (\ref{parpercot}) complete the proof of the theorem. For particular spacetimes, some mention of decoupling was given in \cite{daniel}. It is worth to  emphasize that the above theorem  also works with a source term in the field equations (\ref{fieldeqs}), provided that the mixed projection of the source energy-momentum tensor $ T_{\mu\nu} $ vanishes identically, $ \overline{T}_{* \lambda}= T_{\mu\nu} \xi^{\mu} h_{\;\lambda}^{\nu} \equiv  0 $. We note that  in the case of TMG, decoupling occurs only for a null matter source \cite{df}.

\section{Reduction}

We shall now proceed with the case, when the gradient of the CS scalar field is parallel to the non-null hypersurface orthogonal Killing vector of constant length. That is,
\begin{eqnarray}
\vartheta_{\alpha}& = & \frac{1}{m}\,\xi_{\alpha}\,,~~~~\xi^2= const\,,
\label{varthetaP}
\end{eqnarray}
where $ m $ is a constant parameter.  We  also assume that the cosmological constant vanishes, $ \Lambda=0 $. Then,  the field equations of CS gravity  is given by
\begin{eqnarray}
R_{\mu\nu} + C_{\mu\nu} &= & 0\,,
\label{fieldeqs0}
\end{eqnarray}
It turns out that the following reduction theorem holds:

{\it If a four-dimensional spacetime admits a non-null hypersurface orthogonal Killing vector of constant length and  the gradient of the CS scalar field is parallel to the Killing vector, as given in (\ref{varthetaP}), then CS modified gravity reduces to topologically massive gravity in three dimensions}.

The proof of the theorem is somewhat straightforward. First, we emphasize once again that the identities (\ref{paralcot1}) and (\ref{parperric}) hold independently of the form of  CS scalar field. With this in mind,  we consider the expression
\begin{eqnarray}
R_{**} & = & R_{\mu \nu} \xi^{\mu} \xi^{\nu} = -\frac{1}{2}\,\xi^2 \eta^{\mu}_{\,\,;\mu}\,,
\label{paralric0}
\end{eqnarray}
which can be easily derived using the equation
\begin{eqnarray}
{\xi_{\mu ; \nu}}^{;\nu} + R_{\mu \nu} \xi^{\nu} & = & 0
 \label{xi2oreq}
\end{eqnarray}
and equation (\ref{cdkilling}). Clearly, the expression  in (\ref{paralric0}) vanishes  as
\begin{eqnarray}
 \eta^{\mu} &= & 0
\label{zeroeta}
\end{eqnarray}
for the Killing vector of constant length (see equation (\ref{eta})).
Next, using equations (\ref{cotton4d}) and (\ref{varthetaP})  we obtain that
\begin{eqnarray}
\overline{C}_{*\lambda}&=& \frac{1}{m}\,{\epsilon^{\,\alpha \beta \gamma}}_{(\mu} R_{ \nu) \beta \,;\gamma}\xi_{\alpha} \xi^{\mu}h^{\nu}_{\;\lambda}\,,
\label{parpercot0}
\end{eqnarray}
where we have used the fact that $ \vartheta_{\alpha \beta}= 0 $ for the case under consideration.  It is not difficult to see that this expression  vanishes   as well,
\begin{eqnarray}
\overline{C}_{*\lambda}&=& 0.
\label{parpercot00}
\end{eqnarray}
The remaining step is to examine the projection of equation (\ref{fieldeqs0}) in the direction  orthogonal to the Killing vector of constant length. Taking in the following $ \xi^2=1 $ for certainty,  we appeal to equation (\ref{finalcotperp}), which now takes the form
\begin{eqnarray}
\overline{C}^{\,\lambda \sigma} & = & \frac{\xi_{\tau}}{m}\,
\epsilon^{\,\tau \beta \gamma (\lambda}D_{\gamma}\overline{R}^{\,\sigma)}_{\beta} \,.
\label{finalcotperp0}
\end{eqnarray}
Passing to three-dimensional quantities, by means of equation (\ref{ricci3to4}) and  the fact that
\begin{eqnarray}
\xi_{\tau} \epsilon^{\,\tau \beta \gamma \lambda}=\frac{\varepsilon^{\,\beta \gamma \lambda}}{\sqrt{- ^{(3)}g}}\,
 \,,
\label{finalcotperp0}
\end{eqnarray}
we arrive at the expression for the three-dimensional Cotton tensor
\begin{eqnarray}
^{(3)}C^{\,\lambda \sigma} & = & \frac{1}{m}\,
\epsilon^{\, \beta \gamma (\lambda}D_{\gamma} ^{(3)}{R}^{\,\sigma)}_{\beta} \,,
\label{cot3d}
\end{eqnarray}
where $  \epsilon^{\, \beta \gamma \lambda} $ is the three-dimensional Levi-Civita tensor.  Finally,  with (\ref{ricci3to4}) and (\ref{zeroeta}) in mind, we see that the vacuum field equations of CS modified gravity given in (\ref{fieldeqs0})  reduce to those of TMG \cite{djt}. Thus, we  have
\begin{eqnarray}
^{(3)}R_{\mu \nu } \,+ \,^{(3)} C_{\mu\nu} &= & 0\,.
\label{tmgeqs}
\end{eqnarray}
This completes the proof of the reduction theorem. This theorem  shows that any vacuum solution of TMG uplifted to four dimensions, by adding an extra  flat dimension, will be a solution to CS modified gravity  as well. For instance, the G\"{o}del-type vacuum solution of  TMG found in \cite{percacci} can be extended to four dimensions as a non-trivial (non-general relativity) G\"{o}del solution to CS modified gravity \cite{ah}.

With nonvanishing cosmological constant, $   \Lambda \neq 0 $, a similar reduction to TMG requires an appropriate matter source in the field equation of CS modified gravity. Thus, we must now begin with the equation
\begin{eqnarray}
E_{\mu\nu} + C_{\mu\nu} &= & 8 \pi T_{\mu\nu} \,,
\label{fieldeqsource}
\end{eqnarray}
instead of (\ref{fieldeqs}). Taking the trace of this equation, we have
\begin{eqnarray}
4\Lambda - R &= & 8 \pi T \,.
\label{fieldeqtr}
\end{eqnarray}
Next, using   identity (\ref{paralcot1}) in  (\ref{fieldeqsource}) and taking into account equation (\ref{paralric0}), we find that
\begin{eqnarray}
2 \Lambda - R &= &  16 \pi T_{**} \,.
\label{pareq1}
\end{eqnarray}
With equations (\ref{parperE}) and (\ref{parpercot00}), from (\ref{fieldeqsource}) it also follows that
\begin{eqnarray}
\overline{T}_{* \lambda}= T_{\mu\nu} \xi^{\mu} h_{\;\lambda}^{\nu} \equiv  0
\label{mixpro1}
\end{eqnarray}
Finally, assuming that in three dimensions
\begin{eqnarray}
\overline{T}_{\lambda \sigma}& = & T_{\mu\nu} h^{\mu}_{\;\lambda} h^{\nu}_{\lambda} = 0\,
\label{orthogrE}
\end{eqnarray}
and  taking into account equations (\ref{ricci3to4}) and (\ref{ricscalar3to4}) along with (\ref{zeroeta}), it is easy to see that the orthogonal projection of equation (\ref{fieldeqsource}) results in the field equation of TMG with the cosmological constant \cite{deser}. We have
\begin{eqnarray}
^{(3)}G_{\mu \nu}+ \Lambda h_{\mu\nu}  + \,^{(3)}C_{\mu\nu} &= & 0\,,
\label{tmgeqscos}
\end{eqnarray}
where
\begin{eqnarray}
^{(3)}G_{\mu \nu} & = & ^{(3)}R_{\mu \nu}-\frac{1}{2}\,  ^{(3)}R h_{\mu\nu}\,.
\label{Ein3}
\end{eqnarray}
The trace of this equation,  with (\ref{ricscalar3to4}) in mind, gives
\begin{eqnarray}
^{(3)}R &= & R = 6 \Lambda\,.
\label{trtmgeqscos}
\end{eqnarray}
Comparing this equation with (\ref{fieldeqtr}) and  (\ref{pareq1}) we specify the form of the matter  term  in (\ref{fieldeqsource})
as follows
\begin{eqnarray}
8 \pi T_{**}  &= &  8 \pi T = - 2 \Lambda\,.
\label{matter}
\end{eqnarray}
Thus, with this matter source and the CS scalar field given in (\ref{varthetaP}),  CS modified gravity comprises  all previously known solutions of cosmological TMG (see for instance, Refs. \cite{nutku, gurses} as well as a recent  paper \cite{chow}) when uplifted to four dimensions by adding an extra flat dimension.

\section{Conclusion}

We have examined  the Einstein and Cotton sectors of CS modified gravity in four dimensional spacetimes with a non-null hypersurface  orthogonal Killing vector field, considering all possible projections in directions parallel and orthogonal to the Killing vector. Assuming that CS scalar field is constant along the Killing vector, we have shown that in each case only one of the corresponding  projections (for either Einstein or Cotton sector) vanishes identically. This in turn entails  vanishing of the remaining projections as a consequence of the field equations. Thus, we have proved that  in the model under consideration, the source-free field equations of CS modified gravity decouple into the Einstein and Cotton sectors,  supporting only general relativity solution. We have  emphasized that  such a separation  may  occur in the presence of  a source term as well, provided that the mixed (parallel/orthogonal)  projection of the source energy-momentum tensor vanishes identically.

Next, assuming that the hypersurface orthogonal Killing vector is of constant length and the gradient of the CS scalar field is parallel to the  Killing vector, we have shown that that the vacuum field equations of CS modified gravity  reduce to those of TMG.  We have also extended this result to the case of nonvanishing cosmological constant. In this case, it turned out that  a similar reduction to cosmological TMG requires an appropriate source term  for CS modified gravity.  We have explicitly given the form of the source term that is determined by the cosmological constant.

\section{Acknowledgment}

We thank  S. Deser for interesting  comments.


\begin{thebibliography}{99}

\bibitem{jackiw} R. Jackiw and S. Y. Pi,  Phys. Rev. D {\bf 68} (2003) 104012.
\bibitem{green} M. B. Green and J. H. Schwarz, Phys. Lett. B {\bf 149} (1984) 117.
\bibitem{ay1} S. Alexander and N. Yunes, Phys. Rept. {\bf 480} (2009) 1.
\bibitem{djt} S. Deser, R. Jackiw and S. Templeton,
 Phys. Rev. Lett. {\bf 48} (1982) 975; Ann. Phys. (N.Y.)
{\bf 140} (1982) 372; {\it erratum-ibid}. {\bf 185} (1988) 406.
\bibitem{td} M. Adak and  T. Dereli, arXiv:0807.1832 [gr-qc].
\bibitem{an} A. N. Aliev and Y. Nutku, Class. Quant. Grav. {\bf 13} (1996) L29.
\bibitem{df} S. Deser and  J. Franklin, arXiv:0912.0708 [gr-qc].
\bibitem{wald} R. M. Wald {\em General Relativity},  University of Chicago Press (1984).
\bibitem{carter} B. Carter, J. Math. Phys. {\bf 10} (1969) 70.
\bibitem{daniel} D. Grumiller and N. Yunes, Phys. Rev. D {\bf 77} (2008) 044015.
\bibitem{percacci} R. Percacci, P. Sodano and I. Vuorio, Ann. Phys. (N.Y.) {\bf 176} (1987) 344.
\bibitem{ah} H. Ahmedov and  A. N. Aliev, in preparation (2010).
\bibitem{deser} S. Deser, in {\it Quantum Theory of Gravity}, ed. S. Christensen, A. Hilger Ltd, London (1984).
\bibitem{nutku} Y. Nutku, Class. Quant. Grav. {\bf 10} (1993) 2657.
\bibitem{gurses} M. G\"{u}rses, Class. Quant. Grav. {\bf 11} (1994) 2585.
\bibitem{chow} D. K. Chow, C.N. Pope and E. Sezgin, arXiv:0906.3559 [hep-th].



\end{thebibliography}
\end{document}